\documentclass[%
reprint,
 amsmath,amssymb,
 aps,
]{revtex4-1}

\usepackage{graphicx}% Include figure files
\usepackage{dcolumn}% Align table columns on decimal point
\usepackage{bm}% bold math
\usepackage{relsize}
\usepackage{parskip}
\begin{document}

\preprint{APS/123-QED}

\title{Influence of Discrete Sources on Detonation Propagation in a Burgers Equation Analog System}% Force line breaks with \\

\author{XiaoCheng Mi}

\author{Andrew J. Higgins}%
 \email{Corresponding author: andrew.higgins@mcgill.ca}
\affiliation{%
McGill University, Department of Mechanical Engineering, Montreal, Quebec, Canada
}%

\date{\today}% It is always \today, today,

\begin{abstract}

An analog to the equations of compressible flow that is based on the inviscid Burgers equation is utilized to investigate the effect of spatial discreteness of energy release on the propagation of a detonation wave.  While the traditional Chapman-Jouguet (CJ) treatment of a detonation wave assumes that the energy release of the medium is homogeneous through space, the system examined here consists of sources represented by $\delta$-functions embedded in an otherwise inert medium. The sources are triggered by the passage of the leading shock wave following a delay that is either of fixed period or randomly generated. The solution for wave propagation through a large array ($10^3$-$10^4$) of sources in one dimension can be constructed without the use of a finite difference approximation by tracking the interaction of sawtooth-profiled waves for which an analytic solution is available.  A detonation-like wave results from the interaction of the shock and rarefaction waves generated by the sources. The measurement of the average velocity of the leading shock front for systems of both regular, fixed-period and randomized sources is found to be in close agreement with the velocity of the equivalent CJ detonation in a uniform medium wherein the sources have been spatially homogenized. This result may have implications for the applicability of the CJ criterion to detonations in highly heterogeneous media (e.g., polycrystalline, solid explosives) and unstable detonations with a transient and multidimensional structure (e.g., gaseous detonation waves).

\end{abstract}

\maketitle

%\tableofcontents

\section{\label{sec:I}Introduction}

Detonation waves are supersonic combustion waves in which a shock, propagating through a reactive medium, initiates energy release in the medium that, in turn, supports the initial shock wave. The coupled shock and reaction complex is nearly always observed to propagate at a velocity predicted by the Chapman-Jouguet (CJ) criterion, which states that the flow, upon reaching the end of the reaction zone, moves away from the wave at the local sonic velocity (i.e., Mach number unity) in the wave-fixed reference frame. This criterion is equivalent to stating that the leading edge of the rarefaction wave in the unsteady expansion of detonation products downstream of the wave is equal to the velocity of the detonation wave itself. The CJ criterion provides the necessary condition to close the conservation laws of mass, momentum, and energy and thus solve for the detonation velocity and thermodynamic state of the reacted medium, provided the equation of state is known. The success of this simple criterion, which was formulated more than a century ago \cite{Chapman1899,Jouguet1905}, is intriguing since it is formulated for a wave propagating at a steady velocity into a uniform medium, a scenario that is almost never encountered in experiments examining detonation waves. In gaseous explosives, the structure of the detonation front is always unstable and multidimensional, consisting of detonation “cells” defined by transverse shock waves that propagate across the front.\cite{Lee2008} Similar features are believed to exist in liquid explosives as well.\cite{Fickett2000} These features can be attributed to the fact that detonation waves with reaction zones governed by activated (Arrhenius) kinetics are unstable to perturbations in the transverse direction.\cite{Short1998, Sharpe1997, NgChapter3} In condensed-phase explosives (i.e., polycrystalline high explosives), the reaction zone is usually controlled by the heterogeneity of the medium, wherein shock localization at density inhomogeneities results in local centers of energy release, so-called ``hot spots,'' from which reaction fronts burn-out the rest of the explosive within the reaction zone. Remarkably, despite these multidimensional and unsteady features, the average velocity of detonation fronts is usually very close to the predictions of the CJ criterion, provided that the detonation is not near its propagation limits. In gases, the average velocity of a detonation wave is typically within $1\%$ of the ideal CJ velocity, despite the fact that the local shock front velocity may undergo transient deviations as large as $50\%$ over the cycle of a detonation cell.\\

The usual explanation offered to account for the success of the CJ criterion is that it can still be applied in an average sense to unsteady and multidimensional detonation waves. For example, if a large enough control volume, moving at a steady velocity, can enclose the detonation wave, then the CJ criterion (along with the conservation laws) can still be applied with accuracy to the average flow entering and exiting the control volume. This answer is not entirely satisfying, since the averaging procedure may not be justified. The motivation for the present paper is to put this CJ criterion to a rigorous test by considering an extreme case of a highly discrete media wherein all of the energy release is concentrated in $\delta$-functions in space and time, embedded in an otherwise inert medium. This scenario can be considered the extreme limit of a heterogeneous energetic medium, in contrast with the spatially uniform energetic media considered in the CJ model.\\

A similar examination of reaction-diffusion waves in biological excitable media and flames in energetic media in the limit of spatially discrete sources has recently been undertaken.\cite{GoroshinLeeShoshin1998,Keizer1998,Mitkov1999,Beck2003,Tang2009CTM,Goroshin2011PRE,Tang2011PRE} This approach has generated a number of unique results that would not be encountered in traditional flame theory, such as a limit to flame propagation in adiabatic systems that is distinct from the classical thermodynamic limit.\cite{Tang2009CTM,Goroshin2011PRE,Tang2011PRE} Experimental results showing an independence of flame speed on oxygen concentration for fuel particulates in suspension in a gaseous mixture has provided experimental validation of another prediction of discrete-source flame theory.\cite{Goroshin2011PRE} This approach has resulted in recognition of a separate branch of combustion, i.e.,``discrete combustion,'' that is unique from both traditional homogeneous and heterogeneous combustion.\cite{Mukasyan2008} The present paper is an initial examination into the analogous problem of a discrete-source detonation.\\

Exploring the unsteady or multidimensional detonation dynamics usually necessitates numerical simulations due to the inherent nonlinearity of the governing Euler equations of compressible fluid flow. The ability to numerically resolve spatial and temporal $\delta$-functions is a challenging problem, and thus the examination of simpler, analog systems for which analytic solutions are available is of interest. The present paper utilizes a reactive Burgers equation analog that was originally formulated by Fickett \cite{Fickett1979,Fickett1985Chp,Fickett1985} and Majda \cite{Majda1981}. The advantage of using this analog is its simplicity. Instead of the conservation of mass, momentum, and energy in the Euler equations, the conservation of only one analogous scalar quantity is considered in this analog. This simplification has been rigorously justified by Faria et al. showing that the reactive compressible Navier-Stokes equations can be reduced to a one-dimensional reactive Burgers equation based on an asymptotic analysis.\cite{Faria2014arxiv}\\

This analog approach to examining detonation dynamics has recently generated a variety of noteworthy results, including the ability to capture the period doubling sequence of bifurcations that leads to a chaotically pulsating detonation wave \cite{Radulescu2011PRL,Kasimov2013PRL,Bellerive2014,Faria2014arxiv}, and to qualitatively model the phenomenon of shock-induced ignition in a spatially uniform reactive media governed by two-step induction-reaction kinetics \cite{Tang2013CS}. This approach has been expanded to study non-ideal one-dimensional detonations and was able to capture the instabilities caused by the presence of a loss term that can represent the effects of front curvature or friction on the detonation wave.\cite{Faria2015CS} All of these prior studies treated the medium as being spatially uniform. The present paper is the first to examine discrete-source detonations in analogs based upon the Burgers equation.\\

\section{\label{sec:II}Description of Analog Model}

\subsection{\label{sec:IIA}Burgers Analog Model}

Using the one-dimensional scalar Burgers equation as a model to explore detonation phenomena was proposed by Fickett \cite{Fickett1979,Fickett1985Chp,Fickett1985} and independently by Majda \cite{Majda1981}. For this paper, we will construct our system in an Eulerian reference frame, following Fickett \cite{Fickett1985}, however, the energy released by reaction will be included  as a source term in the conserved variable, as in Majda's model \cite{Majda1981}. This analog system is based on the conservation of $E$, a variable representing some features of an extensive property, such as mass, momentum, or energy, in compressible flow. The volumetric density of $E$, denoted $\rho$, is analogous to an intensive  property, such as density, velocity, or specific energy, in compressible flow. The governing conservation equation is in an Eulerian reference frame, in other words, the conservation law is applied to a specific region of space through which fluid flows, i.e., a control volume. For a reactive flow, $E$ is convected into and out of the control volume by the flow while also being generated by chemical reaction within the control volume. The diffusion of $E$ through the boundary of the control volume is neglected in this model, i.e., an inviscid flow is assumed.\\
\begin{figure}[h]
	\centering
		\includegraphics[width=0.35\textwidth]{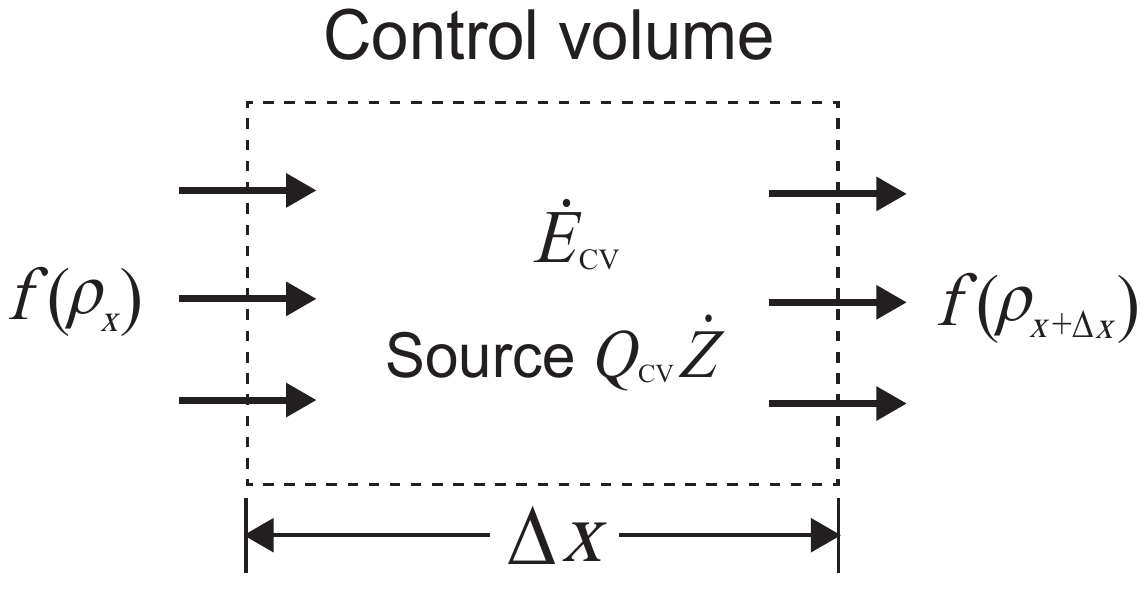}
		\caption{Conservation of $E$ in the control volume}
	\label{Fig3}
\end{figure}

In Fig.~\ref{Fig3}, $f(\rho)$ is the flux of $E$ entering and leaving the control volume, which is analogous to pressure in the momentum equation. $\dot{E}_{\!_\mathrm{CV}}$ is the time rate of change of $E$ in the control volume. The generation rate of energy $E$ within the control volume is equal to the time rate of change of the reaction progress variable, $\dot{Z}$, multiplied by $Q_{\!_\mathrm{CV}}$, the total energy released in the control volume when the reaction is complete. Note that the source energy $Q_{\!_\mathrm{CV}}$ is a property of space only, so it is not advected or compressed in this analog system. In other words, energy sources are assumed to be stationary in a lab-fixed reference frame.\\

The equation of energy conservation for the control volume can be written as follows,
\begin{equation}
\dot{E}_{\!_\mathrm{CV}}=Q_{\!_\mathrm{CV}} \dot{Z} + f\left(\rho_x \right) - f\left(\rho_{x+\Delta x} \right)
\label{Eq1}
\end{equation}
Dividing by $\Delta x$ on both sides of Eq.~\ref{Eq1}, the equation of energy conservation can be obtained in terms of the intensive property, $\rho$,
\begin{equation}
\dot{\rho}=\frac{Q_{\!_\mathrm{CV}}}{\Delta x} \dot{Z} +\frac{f\left(\rho_x \right) - f\left(\rho_{x+\Delta x} \right)}{\Delta x}
\label{Eq2}
\end{equation}
Here, $Z$ represents the mass fraction of product (i.e., $Z=0$ for unreacted and $Z=1$ for fully reacted) in this system.\\

Introducing the spatial density of the energy source, $q=\frac{Q_{\!_\mathrm{CV}}}{\Delta x}$, and taking the limit of Eq.~\ref{Eq2} as $\Delta x$ approaches $0$, the following equation can be obtained,
\begin{equation}
\frac{\partial (\rho - qZ)}{\partial t} + \frac{\partial f(\rho)}{\partial x} = 0
\label{Eq3}
\end{equation}
By introducing $c=\frac{\partial f}{\partial \rho}$, Eq.~\ref{Eq3} can be written as
\begin{equation}
\frac{\partial (\rho - qZ)}{\partial t} + c\frac{\partial \rho}{\partial x} = 0
\label{Eq4}
\end{equation}\\
Equation~\ref{Eq4} is a nonlinear advection equation with reactive source term. Hence, $c$ represents the speed at which $\rho$ is advected. The solution $\rho(x,t)$ is constant along each curve $x-ct=x_{\!_0}$ in the $x$-$t$ plane. These curves are known as the characteristics of the equation, and $c$ is the sound speed relative to a lab-fixed reference frame, which is the slope of the corresponding characteristic in the $x$-$t$ plane.\\

The relation between $f$ and $\rho$ is considered as the equation of state. The function of $f(\rho)$ must satisfy the entropy condition for a shock wave solution, which requires characteristics propagate into the shock in the $x$-$t$ diagram, as time advances.\cite{Leveque1992} For a shock propagating rightwards, the shock speed $S$ satisfies the entropy condition if
\begin{equation}
\left.{\frac{\partial f}{\partial x}} \right|_{\rho_{\!_\mathrm{L}}} = c_{\!_\mathrm{L}} > S > c_{\!_\mathrm{R}} = \left.{\frac{\partial f}{\partial x}} \right|_{\rho_{\!_\mathrm{R}}}
\label{Eq5}
\end{equation}
where subscripts ``R'' and ``L'' indicate the pre-shock and post-shock states, respectively. Eq.~\ref{Eq5} requires $f(\rho)$ to be a convex function, i.e., $\frac{\partial^2 f}{\partial \rho^2}>0$, hence, a quadratic function is chosen as the simplest approximation of the equation of state in the detonation analog \cite{Fickett1979,Fickett1985Chp,Fickett1985,Majda1981}
\begin{equation}
f(\rho)=\frac{1}{2} \rho^2
\label{Eq6}
\end{equation}\\
From Eq.~\ref{Eq6}, a relation between $c$ and $\rho$ can be obtained,
\begin{equation}
\frac{\partial f}{\partial \rho}=c=\rho
\label{Eq7}
\end{equation}
which means that the advection velocity of the local energy density $\rho$ equals to the value of $\rho$ itself. Inserting Eq.~\ref{Eq7} into Eq.~\ref{Eq4}, the one-dimensional inviscid reactive Burgers equation arises,
\begin{equation}
\frac{\partial \rho}{\partial t}+\rho \frac{\partial \rho}{\partial x}=q \frac{\partial Z}{\partial t}
\label{Eq8}
\end{equation}\\

In general, a reaction law is required to specify the dependence of reaction rate, $\frac{\partial Z}{\partial t}$, on $Z$ and $\rho$. However, as the simplest approximation, the assumption of instantaneous reaction, i.e., an infinitely thin reaction zone, is made for the following analysis. Hence, no reaction law needs to be specified.\\

\subsection{\label{sec:IIB}Solution for Spatially Homogeneous Source Energy Release}

In a homogeneous reactive system, where the density of the energy source is uniform, $q$ is a constant. By applying the CJ condition, that the fully reacted flow is exactly sonic relative to the leading shock, i.e., $c_{\!_\mathrm{CJ}}=\rho_{\!_\mathrm{CJ}}=D_{\!_\mathrm{CJ}}$, the CJ detonation velocity for the homogeneous system with density of energy source $q$ can be obtained as\\
\begin{equation}
D_{\!_\mathrm{CJ}}=2q
\label{Eq9}
\end{equation}
The details of this derivation can be found in Appendix~\ref{sec:A}.\\

\begin{figure}[h]
	\centering
		\includegraphics[width=0.4\textwidth]{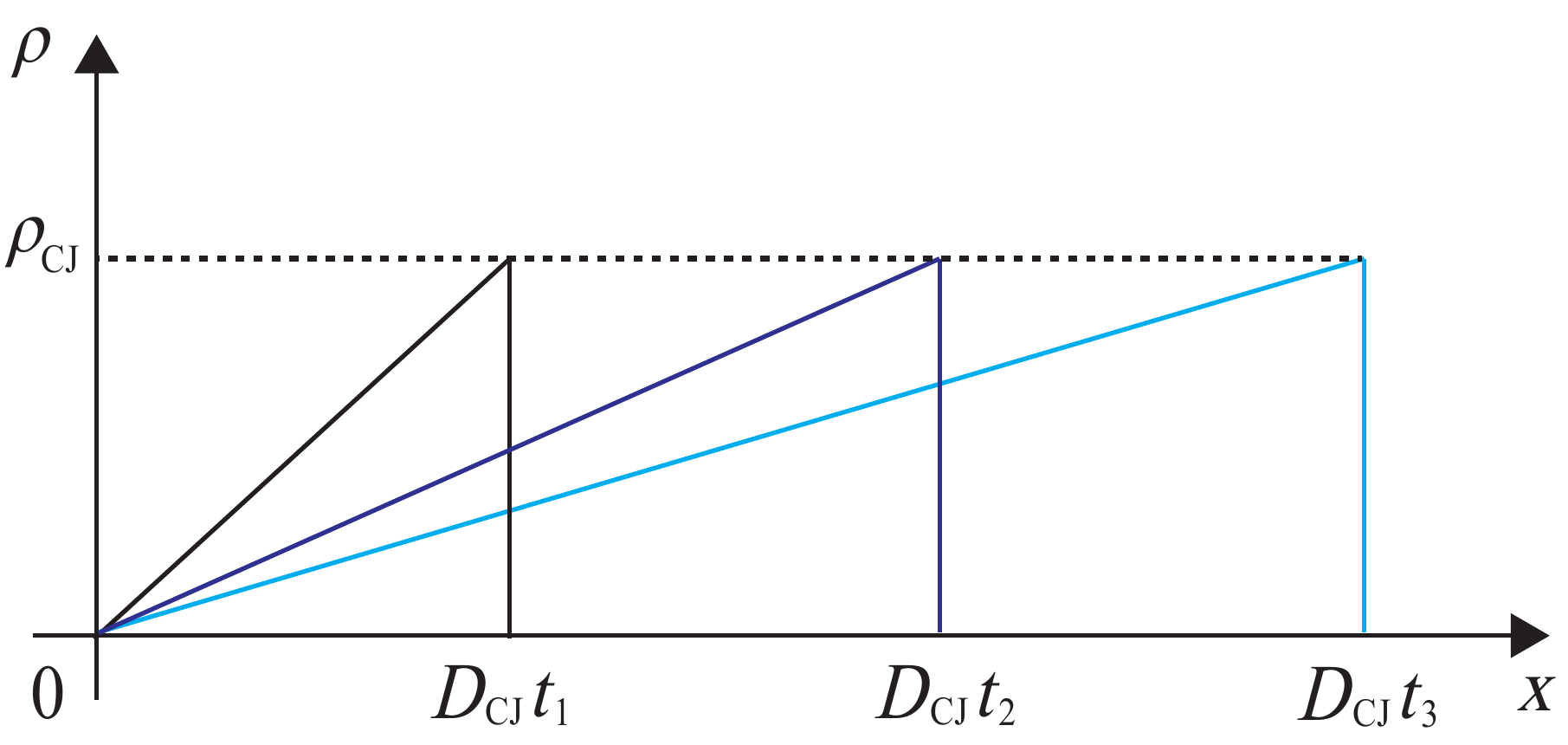}
		\caption{Sample solution for spatially homogeneous source energy release}
	\label{Fig4}
\end{figure}

The solution of $\rho$ for a homogeneous source energy is a right triangular profile over space with a height of $\rho_{\!_\mathrm{CJ}}$ at any time. This triangle gets stretched over time, with its vertical leg propagating to the right at velocity $D_{\!_\mathrm{CJ}}$. The area under the triangular profile of $\rho$ is the total amount of energy released over the distance traveled by the detonation wave at $t$. The solution of $\rho$ for a homogeneous system is illustrated in Fig.~\ref{Fig4}.

\subsection{\label{sec:IIC}Solution for Discrete Source Energy Release}

For a heterogeneous system, the source energy density is not uniform over space. Hence, a spatial distribution of $q$, i.e., $q(x)$, must be specified. In this study, the limit of a highly heterogeneous system is considered wherein the release is concentrated in point-like sources in space. For this case, $q(x)$ can be specified as a summation of equally or randomly spaced $\delta$-functions,
\begin{equation}
q \left(x \right)=\sum_{i}^{N_{\!_{\mathrm{R}}}} \mathrm{\delta} \left(\frac{x-x_{\!_{\mathrm{R,}i}}}{A} \right)
\label{Eq10}
\end{equation}
where $A$ is the total amount of energy released by each discrete source, $x_{\!_{\mathrm{R,}i}}$, is the location of each discrete source, and $N_{\mathrm{total}}$ is the total number of energy sources within an arbitrary length of the reactive system, $L_{\mathrm{total}}$. Hence, given $A$, $N_{\mathrm{total}}$, and $L_{\mathrm{total}}$, the average energy density of a discrete system, $q_{\mathrm{discrete}}$, can be determined. The average energy density is equivalent to the uniform energy density of a homogeneous system,\\
\begin{eqnarray}\nonumber
q_{\mathrm{discrete}} &=& \frac{\mathlarger{\int_{-\infty}^{\infty}} \mathlarger{\sum}_{i}^{{N_{\mathrm{total}}}} \mathrm{\delta} \left(\frac{x-x_{\!_{\mathrm{R,}i}}}{A} \right)\,\mathrm{d}x}{L_{\mathrm{total}}} \\[0.9em]
&=& \frac{A {N_{\mathrm{total}}}}{{L_{\mathrm{total}}}} = \frac{A_{\mathrm{total}}}{{L_{\mathrm{total}}}} = q
\label{Eq11}
\end{eqnarray}
Recall that $\int_{-\infty}^{\infty} \delta (x)\,\mathrm{d}x = 1$. It is important to note that, for a given $A$, $N_\mathrm{R}$, and $L_{\mathrm{total}}$, the discrete systems with both equally and randomly spaced energy sources have the same global energy release as the homogeneous system. However, for a discrete system with equally spaced energy sources, this equivalence can be expressed in a simpler form
\begin{equation}
q_{\mathrm{discrete}}= \frac{A}{L} = q
\label{Eq12}
\end{equation}
where $L$ is the spacing between each two adjacent energy sources. The source can have a finite delay period $\tau$ between being shocked and releasing energy.\\

\begin{figure}[h]
	\centering
		\includegraphics[width=0.4\textwidth]{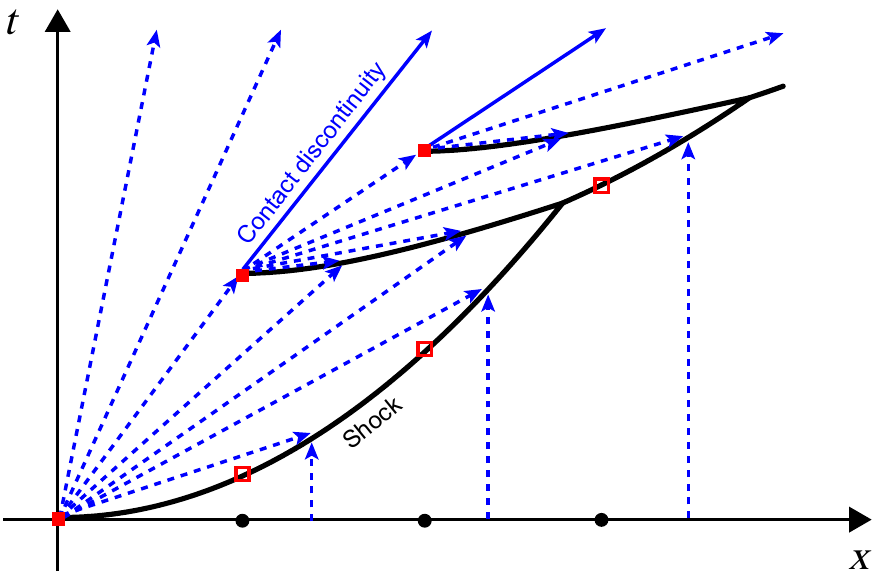}
		\caption{Schematic illustration of the method of characteristics in $x$-$t$ diagram. The right-running characteristics are plotted as dashed, arrowed lines, the trajectories of contact discontinuities as solid, arrowed lines, the trajectories of shocks as thick solid curves, and the locus of sources releasing energy as black solid circles that become red open squares upon being shocked and red solid squares upon releasing their energy.}
	\label{xt_schematic}
\end{figure}

Except for the instants in which the sources are activated, the solution of the discrete system is governed by the non-reactive Burgers equation, 
\begin{equation}
\frac{\partial \rho}{\partial t}+\rho \frac{\partial \rho}{\partial x}= 0
\label{Eq13}
\end{equation}
As this system has only right-running characteristics, the solution of a shock propagating with a linear profile of $\rho$ upstream and downstream of the wave can be treated analytically using the method of characteristics. The detailed solution for the propagating shock can be found in Appendix~\ref{sec:B}. The shock-shock and shock-contact interactions can be similarly treated analytically. A schematic $x$-$t$ diagram of how the method of characteristics is used to solve for the trajectory of the shock waves, contact discontinuities, and their subsequent interactions is shown in Fig.~\ref{xt_schematic}.

\section{\label{sec:III}Implementation of Analog Model}

\begin{figure}[h]
	\centering
		\includegraphics[width=0.35\textwidth]{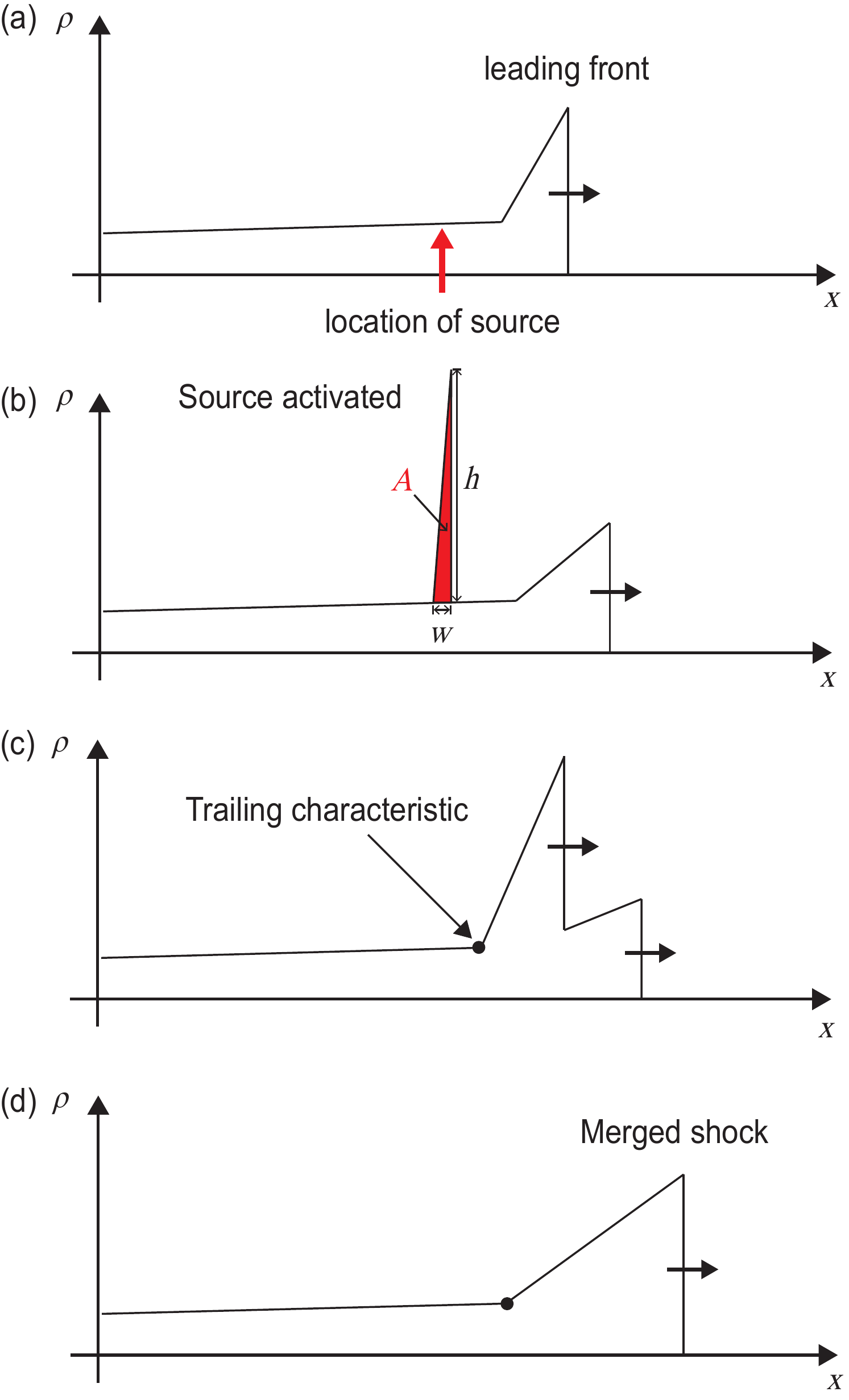}
		\caption{Illustration of the implementation of source activation: (a) source during delay period after being shocked, (b) source energy release inserted as a triangular profile to approximate $\delta$-function, and (c) shortly after being activated, and (d) the second shock now caught up and merged with the leading shock.}
	\label{Fig5}
\end{figure}

\begin{figure}[h]
	\centering
		\includegraphics[width=0.35\textwidth]{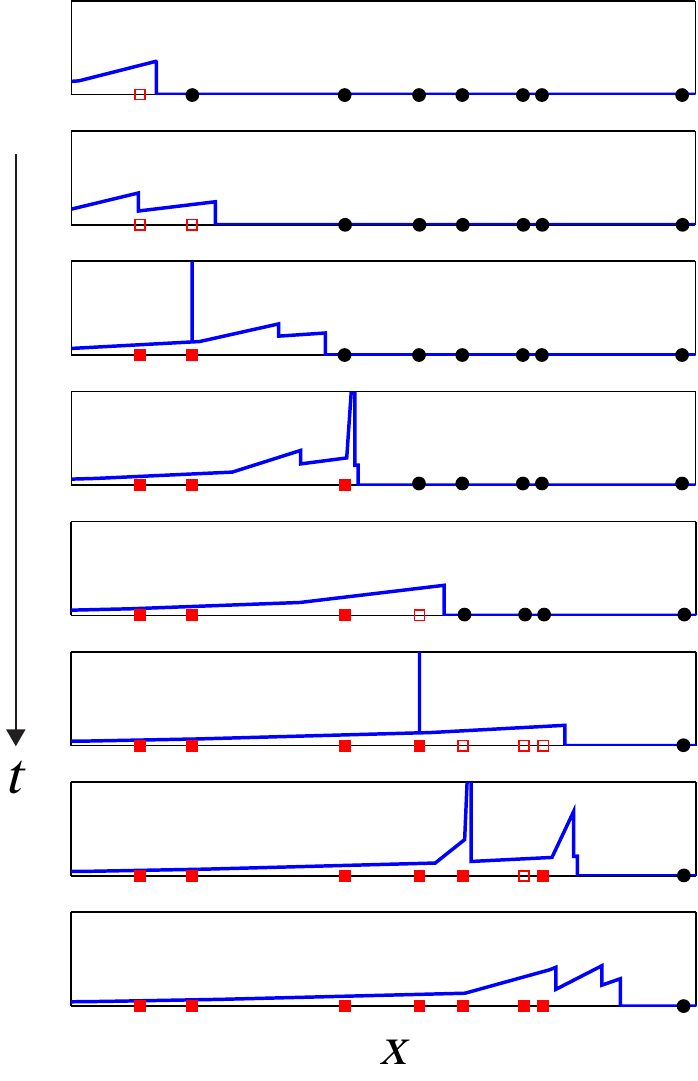}
		\caption{Sample solutions of $\rho$ for randomly spaced energy sources with random period delay. Circles are unshocked sources, empty squares shocked sources during delay period, and solid squares sources with energy released.}
	\label{sample_solution}
\end{figure}

The methodology of constructing a solution to the problem of a shock propagating in a system consisting of discrete sources is implemented via a computer program that tracks the trajectory of the leading shock, internal shocks, contact discontinuities, and the slopes of the sawtooth regions between these various elements. Although a computer is used to track the interacting elements, we emphasize that the solution is entirely analytic and no finite difference approximation is made. When the leading shock has crossed a source location and the prescribed delay period has elapsed, the solution is paused and the new source inserted into the profile of $\rho$ at the source location. To implement this, the $\delta$-function, with an area of $A$ under it, is approximated as a triangular profile with very small width $w$ and height $h=\frac{2A}{w}$, as illustrated in Fig.~\ref{Fig5}. Sample solutions of $\rho$ for randomly discrete systems are shown in Fig.~\ref{sample_solution}\\

The only convergence parameters in this method are the source width $w$ and the precision with which the time of element interactions (shock-shock, shock-contact) are solved for via a numerical root-finding algorithm. It is relatively easy to verify that the solution is independent of these parameters once they are made sufficiently small (width of $\delta$-function $w/L = 10^{\mathord{-} 10}$ and convergence factor in solving for time of wave intersections $\epsilon/t_c = 10^{\mathord{-} 4}$, where $t_\mathrm{c} = L/D_{\!_\mathrm{CJ}}$ is the characteristic time for this study). This convergence study was performed in the range of $w/L = 10^{\mathord{-} 9}$-$10^{\mathord{-} 12}$ and $\epsilon/t_\mathrm{c} = 10^{\mathord{-} 3}$-$10^{\mathord{-} 5}$ for all results reported in this paper, and the results were found to be independent of the value of these parameters.\\

\section{\label{sec:IV}Results}

In the results reported below, the spatial coordinate $x$ is normalized by the average spacing between each two adjacent sources $L$, leading shock velocity $D$ by $D_{\!_\mathrm{CJ}}$, and time $t$ by $t_\mathrm{c}$. Thus, the dimensionless results are only dependent upon $q$ and $\tau$.

\subsection{\label{sec:IVA0}Equally Spaced Energy Sources with Zero Delay}

For the case of zero delay, i.e., a source releases its energy immediately upon contact with the leading shock front, it is possible to solve analytically for the average propagation velocity without having to rely upon a computer program to track the wave interactions. In this case, the $\delta$-function is inserted at the front and there is no interaction between the resulting blast wave and the pre-existing sawtooth profile. In this case, the decay of the shock front is given by \cite{Whitham1974}
\begin{equation}
x_{\mathrm{s}}=\sqrt{2At}
\label{Eq14}
\end{equation}
From Eq.~\ref{Eq14}, the time required for the shock front to reach the next source a distance $L$ away can be solved
\begin{equation}
\left. t \right|_{x_{\mathrm{s}}=L} = \frac{L^2}{2A}
\label{Eq15}
\end{equation}
Hence, the average velocity for the shock front to travel to the next source over a distance $L$ is
\begin{equation}
v_{\mathrm{avg}} = \frac{L}{\left. t \right|_{x_{\mathrm{s}}=L}} = \frac{2A}{L} = 2q = D_{\!_\mathrm{CJ}}
\label{Eq16}
\end{equation}
which is equal to the CJ detonation velocity of the equivalent homogeneous media. The detailed derivation of $D_{\!_\mathrm{CJ}}=2q$ can be found in Appendix~\ref{sec:A}. Since there is no influence from one cycle to the next, the wave will continue to propagate in a periodic manner, with the average wave speed always equal to this value.

\subsection{\label{sec:IVA}Equally Spaced Energy Sources with Fixed Period Delay}

\begin{figure}[h]
	\centering
		\includegraphics[width=0.45\textwidth]{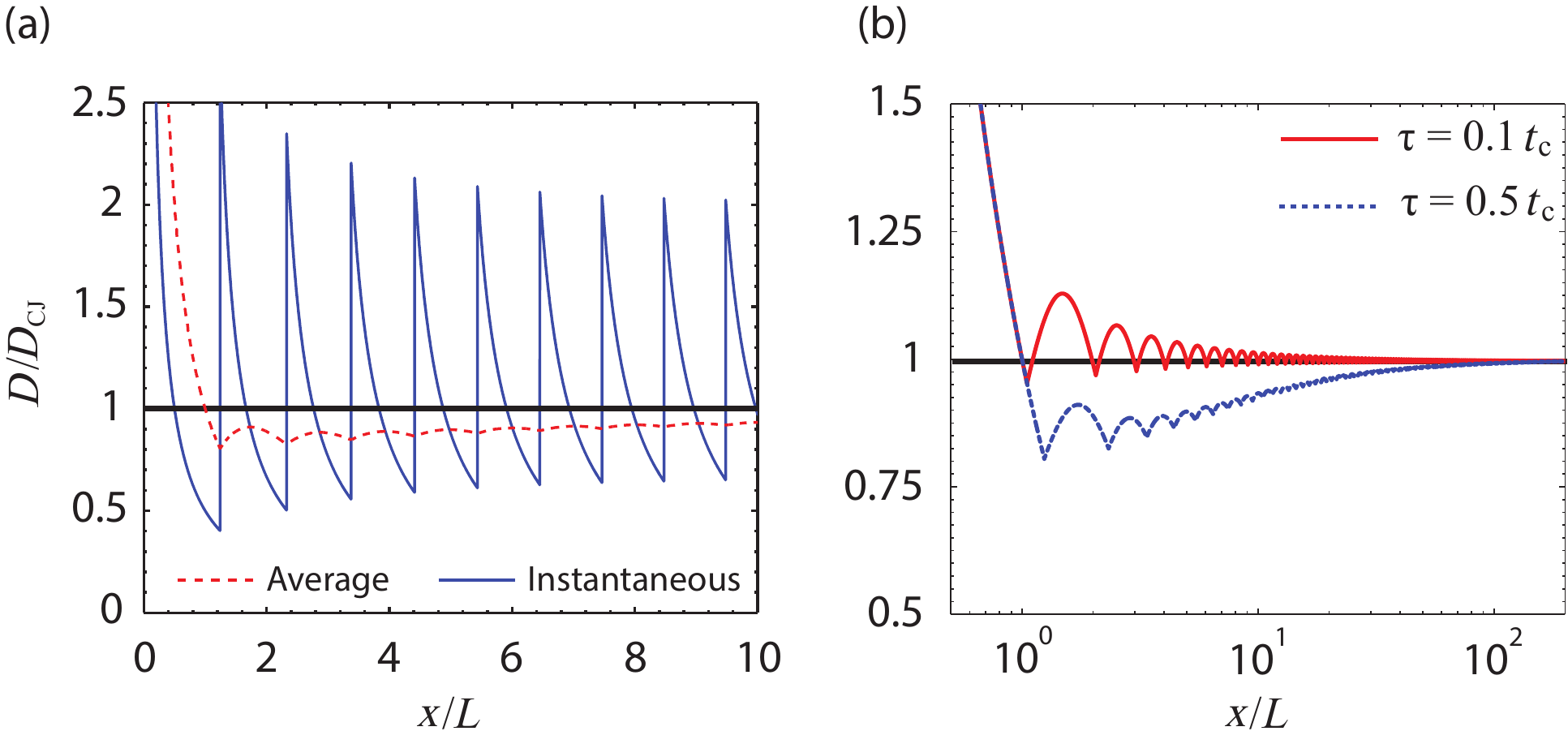}
		\caption{Velocity history of a system where sources were equally spaced with an average energy density of $q=0.5$ and (a) a fixed delay of $\tau=0.5 t_\mathrm{c}$ over a short distance in linear scale, and (b) fixed delays of $\tau=0.5 t_\mathrm{c}$ and $\tau=0.1 t_\mathrm{c}$ over a long distance in logarithmic scale.}
	\label{Fig6}
\end{figure}

The results shown in Fig.~\ref{Fig6} are of a case where discrete sources were equally spaced along the $x$-axis with an average energy density of $q=0.5$ (i.e., $D_{\!_\mathrm{CJ}}=1$) and fixed delays of $\tau=0.5 t_\mathrm{c}$ and $\tau=0.1 t_\mathrm{c}$.\\

In Fig.~\ref{Fig6}(a), the history of the instantaneous and average leading shock velocities for $\tau=0.5 t_\mathrm{c}$ are compared to the CJ velocity of the system. The instantaneous velocity was obtained analytically from the solution, while the average velocity was computed by simply dividing the distance the leading shock had traveled by the time elapsed.\\

As shown in Fig.~\ref{Fig6}(a), the instantaneous velocity fluctuated between values higher and lower than the CJ velocity, while the average velocity slowly approached the CJ velocity. This fluctuation appeared to be in self-repetitive cycles as the average velocity approached the CJ value. As shown Fig.~\ref{Fig6}(b), for a short delay time, $\tau=0.1 t_\mathrm{c}$, the wave is overdriven early in the propagation. As the leading shock traveled a sufficiently long distance ($x/L = 200$), the average velocities for both cases of $\tau=0.5 t_\mathrm{c}$ and $\tau=0.1 t_\mathrm{c}$ converge to the CJ velocity with a difference less than $2.5\%$ of $D_{\!_\mathrm{CJ}}$.\\

\begin{figure}[h]
	\centering
		\includegraphics[width=0.45\textwidth]{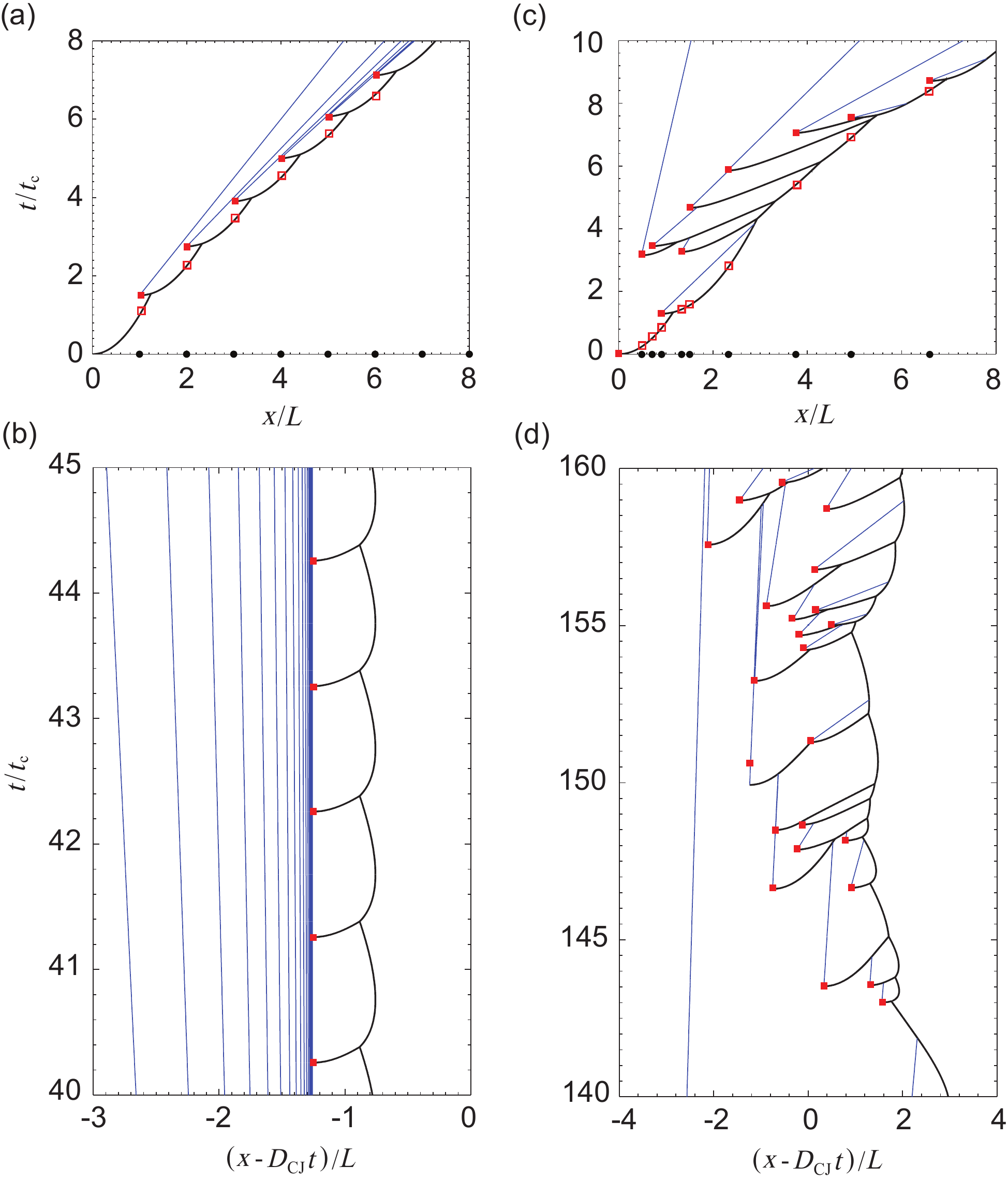}
		\caption{$x$-$t$ diagrams of wave processes, (a) in a lab-fixed reference frame and (b) in a wave-attached frame moving at the CJ velocity, of a system where sources were equally spaced with an average energy density of $q=0.5$ and a fixed delay of $\tau=0.5 t_\mathrm{c}$. $x$-$t$ diagrams of wave processes, (c) in a lab-fixed reference frame and (d) in a wave-attached frame moving at the CJ velocity, of a system where sources were randomly spaced with an average energy density of $q=0.5$ and random delay in the range $0.1$ to $1 t_{\mathrm{c}}$. The trajectories of contact discontinuities are plotted as blue lines, the trajectories of shocks as black curves, and the locus of sources releasing energy as black solid circles that become red open squares upon being shocked and red solid squares upon releasing their energy.}
\label{Fig7}
\end{figure}

To better examine the interaction of waves and source energy release during the propagation of a detonation wave, the location and time of source activation, and the trajectories of shocks and contact surfaces were plotted in the $x$-$t$ diagrams shown in Fig.~\ref{Fig7}(a) and (b).\\

Figure~\ref{Fig7}(a) shows the $x$-$t$ diagram constructed in a lab-fixed reference frame. A shock and a contact surface originate each time a new source releases its energy. The characteristics are not shown in this figure for clarity, but they have a linear profile between the shocks and contact surfaces, as shown in the schematic of Fig.~\ref{xt_schematic}. The shock generated by a newly released source eventually catches up and accelerates the leading shock. It is interesting to note that, as the detonation propagates, the trajectories of the contact surfaces, i.e., the trailing characteristics of the rarefaction waves, tend to coalesce along the path in $x$-$t$ space where source energy release events are occurring.\\
\begin{figure*}[t]
	\centering
		\includegraphics[width=1.0\textwidth]{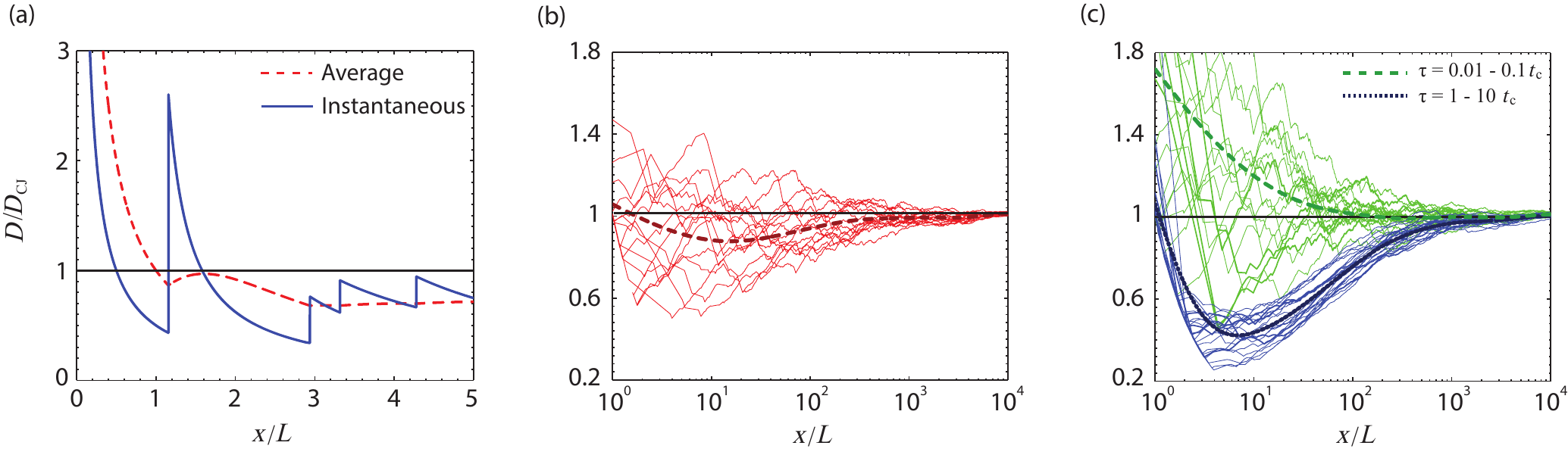}
		\caption{(a) Velocity history of a system where sources were randomly spaced with an average energy density of $q=0.5$ and random time delays over the range $0.1$ to $1 t_\mathrm{c}$ over a short distance in linear scale. (b) The average wave velocity for a set of $20$ realizations of a system of randomly spaced sources with an average energy density of $q=0.5$ and random time delays over the range $0.1$ to $1 t_{\mathrm{c}}$ over a long distance in logarithmic scale. (c) The average wave velocity for two sets of $20$ realizations of systems of randomly spaced sources with an average energy density of $q=0.5$ and random time delays over the ranges $0.01$ to $0.1 t_{\mathrm{c}}$ and $1$ to $10 t_{\mathrm{c}}$ over a long distance in logarithmic scale. The thick dashed lines in (b) and (c) are the ensemble averaged velocities as a function of $x/L$.}
	\label{Fig8}
\end{figure*}

Figure~\ref{Fig7}(b) shows the $x$-$t$ diagram constructed in a reference frame moving at the CJ velocity of a detonation in the equivalent homogeneous system. The time interval ($t=40$ to $45 t_\mathrm{c}$) shown in Fig.~\ref{Fig7}(b) is after the leading shock has traveled a sufficiently long distance so the average velocity is very close to the CJ value. The leading shock front periodically oscillates around approximately the same location relative to the reference frame moving at $D_{\!_\mathrm{CJ}}$. The sources release their energy at almost equal time intervals. As can be clearly observed in this figure, the trailing characteristics associated with prior sources pass extremely close to the locus of new sources.

\subsection{\label{sec:IVB}Randomly Spaced Energy Sources with Random Period Delay}

The locations of energy sources are distributed within a distance $L_\mathrm{total}$ using a random number generator, i.e., $x_{\!_{\mathrm{R},i}} \sim \mathrm{U} [0, L_\mathrm{total}]$, and the delay time associated with each source is randomly generated within a prescribed range, i.e., $\tau_{i} \sim \mathrm{U} [\tau_\mathrm{min}, \tau_\mathrm{max}]$. The results shown in this section are for the case where discrete sources of energy were randomly spaced with an average energy density of $q=0.5$. In this case, $L$ is the average spacing between sources, $L=\frac{q L_\mathrm{total}}{A}$.\\

Figure~\ref{Fig8}(b) shows the average wave velocity as a function of propagation distance for a set of $20$ realizations of a system with randomly spaced sources and random time delays over the range $0.1$ to $1 t_{\mathrm{c}}$. While the average velocities exhibit wide fluctuations, particularly early in the propagation, the average velocities all approach the equivalent CJ velocity of the homogenized system. The fact that many of the individual realizations exhibit very large velocities early on is attributed to the fact that random placement of sources makes it probable that two sources will be located quite close together, resulting in an initially highly overdriven wave. A thick dashed line, corresponding to an ensemble average of $20$ such solutions, is also shown to assist the reader in identifying the overall trend. In Fig.~\ref{Fig8}(c), similar ensembles are shown for the cases of random delay times over the interval $0.01$ to $0.1 t_{\mathrm{c}}$ and the interval $1$ to $10 t_{\mathrm{c}}$. In the case of the very short delay, the ensemble averaged velocity decays monotonically to the CJ velocity of the homogenized system. For the case with long delay time, most of the individual cases decay to well below the equivalent CJ velocity due to the time required for the shock waves generated by the sources to catch up to the leading shock. Upon encountering hundreds of subsequence sources, however, the average velocity exhibits a steadily increasing value, asymptotically approaching the CJ velocity. In all cases, after propagating through $10^4$ sources, the average wave velocity is within $1\%$ of the CJ velocity.\\

The wave processes of this system with randomly spaced sources and random delay are plotted in the $x$-$t$ diagram shown in Fig.~\ref{Fig7}(c) and (d). The $x$-$t$ diagrams constructed in both a lab-fixed reference frame and a reference frame moving at $D_{\!_\mathrm{CJ}}$ are shown. Again, the characteristics in the regions of constant slope between the various shocks and contact surfaces are not plotted in these figures for clarity. No self-repetitive wave interactions can be observed even after the wave front had traveled a significantly long distance. The phenomenon of the trailing characteristics of rarefaction waves coinciding the loci of source firings that was observed with a regular spacing of sources (Section~\ref{sec:IVA}) is not observed in the case with random sources.\\

\section{\label{sec:V}Discussion}

When the release of the conserved variable (here, representing energy) is concentrated in discrete, point-like sources and then released when initiated by a passing shock wave in the system described by the Burgers equation considered in this paper, the resulting blast waves interact and merge with each other and with the leading shock.  While the instantaneous velocity of this front exhibits large variations, the average velocity of the front after traversing many hundreds or thousands of sources is very close (within $2.5\%$) of the ideal Chapman Jouguet velocity of the equivalent uniform medium in which the sources have been homogenized throughout the medium. This result was observed both with regularly spaced sources with a fixed delay time between the shock and with systems with randomly distributed sources and random delay time, as can be seen in Figs.~\ref{Fig6} and \ref{Fig8}, respectively. This result provides a strong validation of the CJ criterion for highly spatially heterogeneous reactive media, at least within the Burgers equation analog system. It may also provide some rationale as to why the CJ criterion is so successful in describing detonations in gases which are characterized by an unstable, multidimensional cellular structure. If this result, namely that a detonation will always propagate on average at the velocity of the CJ detonation in an equivalent homogeneous system, is proven to be universal, it may even supplant the traditional CJ criterion, which was formulated for detonation waves propagating at steady velocity in uniform media.\\

When a source of the conserved variable is released as a $\delta$-function, it evolves into a triangular, sawtooth profile with a leading shock followed by a linear rarefaction. In the present study, the source release was initialized as a triangular profile, but it can be shown that any non-negative profile with a local maximum (i.e., single hump) will eventually develop into this forward propagating triangular (or  “sawtooth”) profile wave \cite{Whitham1974}. Thus, we do not believe our results are influenced by the selection of a triangular profile in introducing the sources. The arrival of the shock at the leading front, and the subsequent merging of these shocks, accelerates and sustains the leading front, while the following rarefaction results in a continued deceleration of the wave. For the case of regularly spaced sources with a fixed delay period, the time at which the sources release their energy evolves to become very close to the time of arrival of the locus of trailing rarefactions from all prior sources, as seen in Fig.~\ref{Fig7}. This phenomenon, in which the trailing rarefactions appear to pile-up on top of each other and become coincident with the time at which the new sources are released, resulted in considerable numerical difficulty in implementing the event-driven solution technique used in this paper, since it was necessary to determine the intersection point in space and time of all interacting elements (shocks, trailing rarefactions, and new sources) in order to ascertain which interaction occurs next.  This condition may also be the discrete-source analog to the traditional CJ criterion, wherein the leading edge of the trailing rarefaction of the unsteady expansion of detonation products becomes coincident with the sonic locus at the end of the reaction zone.\\

The treatment of detonation wave dynamics by considering a detonation as an ensemble of interacting waves originating from discrete sources may have application beyond just computing the average velocity of detonation propagation in an effectively infinite medium. When losses are present (e.g., lateral expansion due to yielding confinement, momentum losses due to boundary layers in narrow channels, etc.), detonations no longer propagate at the ideal CJ detonation velocity even in homogeneous media, and calculation of the correct propagation velocity necessitates solving for the reaction zone structure with losses present. The calculation procedure consists of solving an eigenvalue problem with the post-shock state forming one boundary and a reaction zone structure that passes smoothly through a sonic point (i.e., saddle condition) defining the other boundary. While this technique is well defined for a smooth, laminar-like wave structure in a homogeneous medium (see Ref.~\cite{Li2015POF}), the procedure for heterogeneous or multiphase detonations has yet to be formulated, and must be treated via direct numerical simulation. Results in recent years examining detonations in condensed-phase explosives with large-scale heterogeneities have generated results that cannot be explained using conventional front curvature theory, which assumes a laminar-like wave structure. In the presence of losses, a system of interacting, multidimensional blast waves originating from spatially discrete sources may give rise to a percolating-like phenomenon in which waves would be able to propagate from source to source under conditions wherein a wave in the equivalent homogeneous medium would fail. This type of behavior has been recently demonstrated in detonation simulations using the reactive Euler equations with a medium in which a spatial inhomogeneity has been introduced.\cite{Li2014CS} This approach to detonation dynamics may provide a theoretical framework to account for the anomalous results in highly heterogeneous explosives.\cite{Higgins2014APS}\\

Since the resulting wave structure in the system studied in this paper is seen to be an interacting ensemble of decaying sawtooth-profile blast waves described by the classical inviscid Burgers equation, it would be of interest to further explore these results within the framework of Burgers turbulence. In recent years, there has been a resurgence of examination of the Burgers equation to study scaling exponents in the power law of the probability distribution as solutions of the equation, initialized with white noise for example, decay to a sawtooth profile (see thorough review by Bec and Khanin \cite{Bec2007Burgers}). The fact that turbulence-like scaling is obtained from the inviscid, scalar Burgers equation, wherein all viscous dissipation occurs in the discontinuities embedded within the solution, is a remarkable finding. Some of the results of the present study, in which randomly spaced sources were triggered with random delay times, could be considered in the class of “kicked Burgers turbulence,” but under a scenario wherein the forcing function is triggered or synchronized with the passage of the shock front.

\section{\label{sec:VI}Conclusion}

Solutions to the inviscid Burgers equation with spatially and temporally discrete sources of the conserved variable are released when activated by passage of the leading shock front were examined. The governing equation was non-reactive outside of the instants in time when the release of the sources occurred, enabling the solution to be constructed entirely analytically. The release of a source was treated by inserting a $\delta$-function into the solution, and then allowing the resulting waves to interact. The subsequent interaction of the blast waves in the Burgers equation was observed to propagate, on average, at a speed identical to the Chapman-Jouguet speed of a steady detonation in an equivalent homogeneous system with the same average spatial release of the conserved variable. This result applied to both systems with uniformly spaced sources and fixed periods of delay between shock and release as well as systems with randomized energy release (i.e., sources randomly spaced and with random delay time, but with the sources still activated by the passage of a shock). In the regular spacing and delay case, as the solution progressed, the locus of trailing rarefaction waves evolved to become coincident with the release of new sources, a result that may have some connection to the classical CJ criterion for homogenous media. In a randomly spaced system with random delays, the release of the sources and rarefactions from prior sources did not exhibit this coincident behavior, but the resulting wave complex still propagated with an average velocity equal to that predicted by the CJ criterion of the equivalent homogeneous medium.

\appendix

\section{\label{sec:A}Chapman-Jouguet Detonation Velocity for a System with Spatially Homogeneous Energy Release}

\begin{figure}[h]
	\centering
		\includegraphics[width=0.35\textwidth]{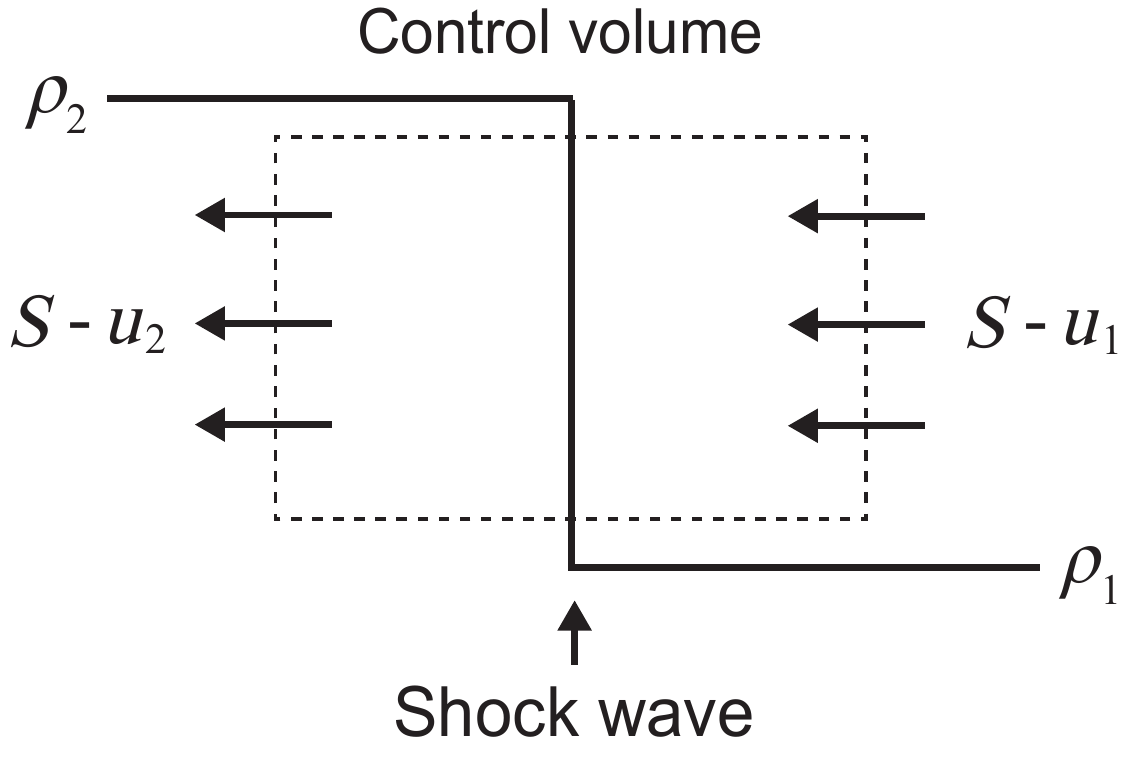}
		\caption{Schematic representation of the control volume enclosing a nonreactive shock wave.}
	\label{FigA1}
\end{figure}

For a control volume enclosing and attached to a nonreactive shock wave (as illustrated in Fig.~\ref{FigA1}), the flux of $E$ leaving the control volume must equal to that entering the control volume
\begin{equation}
\left(S-u_2 \right)\rho_2-\left(S-u_1 \right)\rho_1 = 0
\label{EqA1}
\end{equation}
where $S$ is the speed of shock propagation, and $u$ is the flow speed with respect to a lab-fixed reference frame. Since $f(\rho)=u\rho$, $S$ can be derived from Eq.~\ref{EqA1} as
\begin{equation}
S= \frac{f \left( \rho_2 \right) - f \left( \rho_1 \right)}{\rho_2 - \rho_1}
\label{EqA2}
\end{equation}
which is known as the Rankine-Hugoniot jump condition.\\

Considering the specific equation of state as Eq.~\ref{Eq6}, $S$ can be solved as
\begin{equation}
S=\frac{1}{2} \left(\rho_1 + \rho_2 \right)
\label{EqA3}
\end{equation}
This is the well-known result that a shock, as a solution of the Burgers equation, propagates at the average of the pre-shock and post-shock characteristic speed.\\

For a detonation wave, the difference between the fluxes leaving and entering the control volume equals the rate of $E$ being released or absorbed by reaction within the control volume. Since the energy source is assumed to be stationary in a lab-fixed reference frame, it enters and leaves the control volume at the speed of detonation propagation, $D$. Thus, the conservation equation across a detonation wave can be written as
\begin{equation}
(D-u_2)\rho_2-(D-u_1)\rho_1 = D \left( Z_2 - Z_1 \right) q
\label{EqA4}
\end{equation}
Note that $Z_1$ is always equal to $0$ since the reactant in front of the detonation wave is fully unreacted. By $f(\rho)=u\rho$, $D$ can be expressed as
\begin{equation}
D = \frac{f \left( \rho_2 \right) - f \left( \rho_1 \right)}{\left(\rho_2 - q Z_2 \right) - \rho_1}
\label{EqA5}
\end{equation}
For any value of $Z_2$ such that $0 < Z_2 \leq 1$, Eq.~\ref{EqA5} can be plotted as a straight line with slope $D$ in $\rho$-$f(\rho)$ space, which is analogous to the Rayleigh line for the governing equations of compressible fluid flow. The analog of the Hugoniot curve is Eq.~\ref{Eq6}, the equation of state.\\

The Chapman-Jouguet condition, which states that the flow becomes sonic relative to the detonation, i.e., $c_{\!_\mathrm{CJ}}=\rho_{\!_\mathrm{CJ}}=D_{\!_\mathrm{CJ}}$, when the material is fully reacted, i.e., $Z_2=1$, requires the following relations to be satisfied simultaneously,
\begin{equation}
 \left \{  
\begin{array}{l l}
    \rho_{\!_\mathrm{CJ}} = D_{\!_\mathrm{CJ}}\\[0.8em]
    D_{\!_\mathrm{CJ}} = \frac{1}{2} \left( \frac{{\rho_{\!_\mathrm{CJ}}}^2}{\rho_{\!_\mathrm{CJ}}-q} \right)
  \end{array} \right.
\label{EqA6}
\end{equation}
Here, $\rho_1=0$ as the leading shock wave is assumed to be very strong, and the equation of state has the form of Eq.~\ref{Eq6}. By solving Eq.~\ref{EqA6}, the CJ detonation speed for a homogeneous system with energy density $q$ can be obtained as
\begin{equation}
D_{\!_\mathrm{CJ}}=2q
\label{EqA7}
\end{equation}

\begin{figure}[h]
	\centering
		\includegraphics[width=0.4\textwidth]{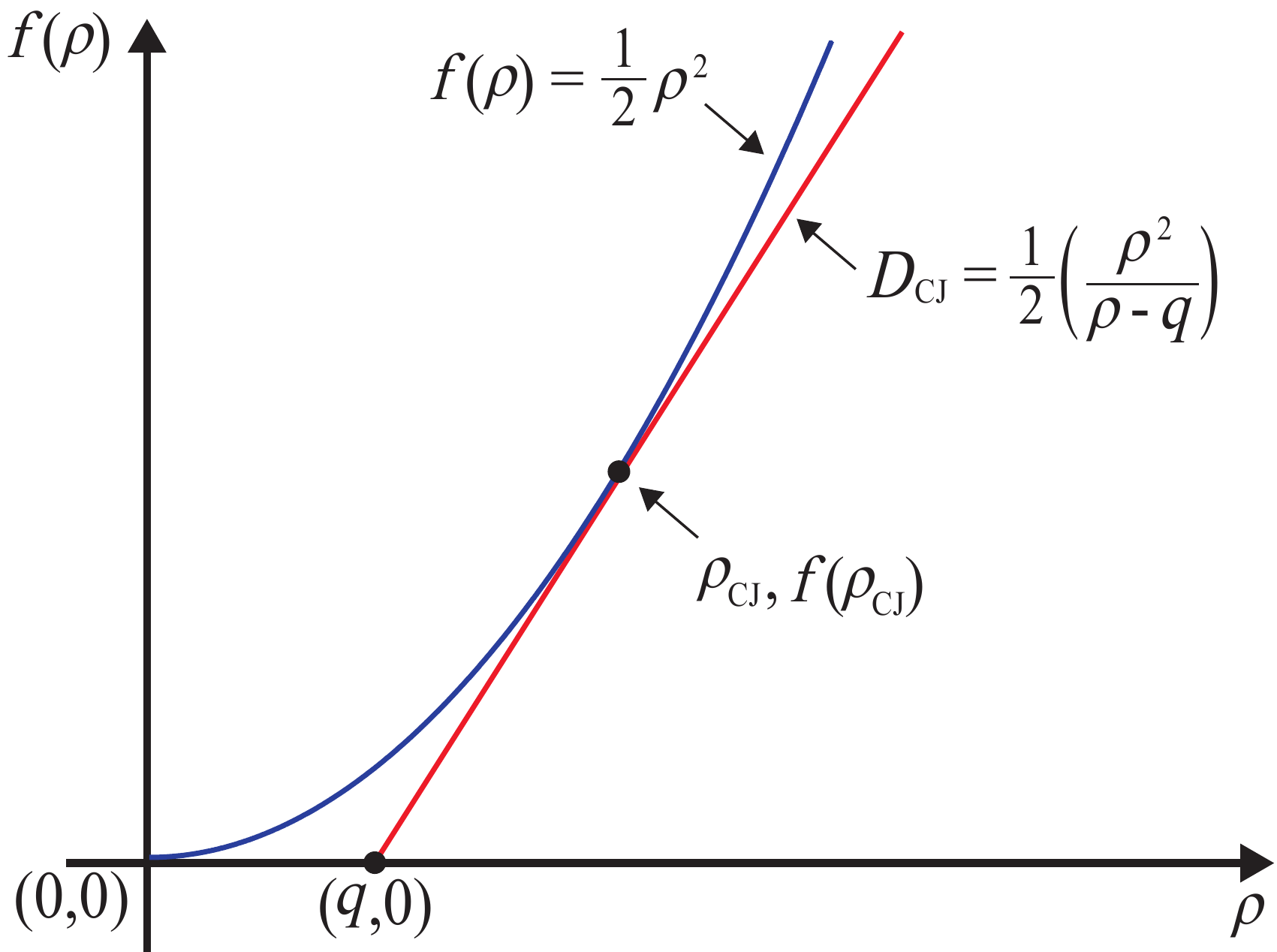}
		\caption{The analog of the Chapman-Jouguet condition shown in $\rho$-$f(\rho)$ diagram.}
	\label{FigA2}
\end{figure}

As shown in Fig.~\ref{FigA2}, $D_{\!_\mathrm{CJ}}$ can be equivalently determined by finding the slope $D$ corresponding to the Rayleigh line with $Z_2=1$ tangent to the Hugoniot curve $f(\rho)=\frac{1}{2} \rho^2$ in $\rho$-$f(\rho)$ space.

\section{\label{sec:B}Propagation of a Shock Wave Solved via the Method of Characteristics}

\begin{figure}[h]
	\centering
		\includegraphics[width=0.4\textwidth]{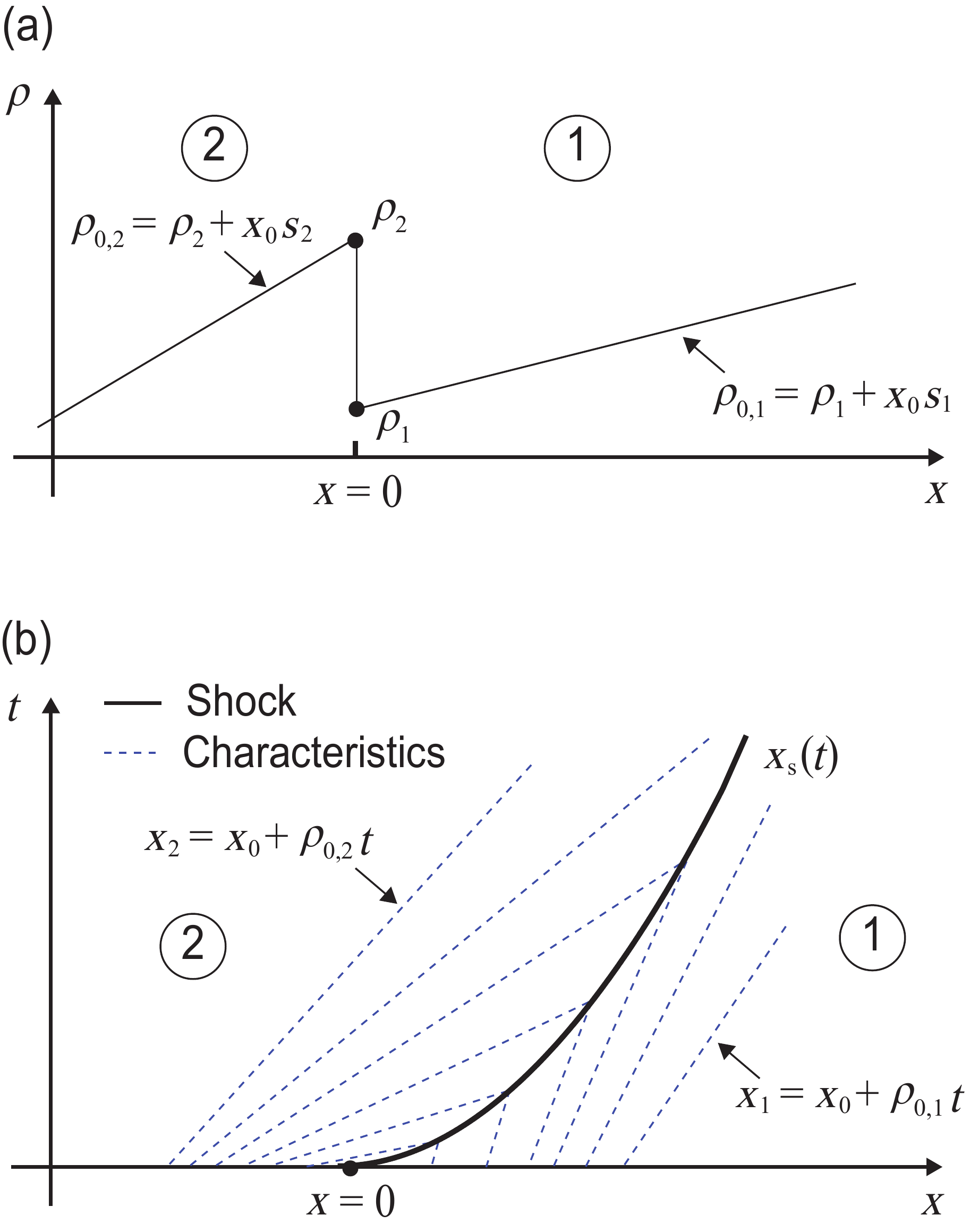}
		\caption{Illustration of (a) a shock wave located between two linear profiles of $\rho$ upstream and downstream of the shock and (b) the trajectory of the shock wave solved using the method of characteristics in an $x$-$t$ diagram.}
	\label{FigB1}
\end{figure}

In the system with energy sources consisting $\delta$-functions in space, the shock waves always propagate with a linear profile of $\rho$ upstream and downstream of the shocks. For a special case of the leading shock with a zero-slope profile of $\rho$ upstream, the position of the shock front as a function time can be solved as Eq.~\ref{Eq14}.\cite{Whitham1974}. Figure~\ref{FigB1}(a) is a snapshot of a shock located between two linear profiles of $\rho$ at an arbitrary time. For simplicity, we set the location of the shock to be $x=0$ at this instant $t=0$. The values of $\rho$ immediately in front and behind the shock are $\rho_1$ and $\rho_2$, respectively. The upstream and downstream linear profiles of $\rho$ at $t=0$, i.e., $\rho_{0,1}$ and $\rho_{0,2}$ can be expressed as follows,
\begin{equation}
\rho_{0,1}=\rho_1+x_0 s_1 \;\;\; \rho_{0,2}=\rho_2+x_0 s_2
\label{EqB1}
\end{equation}
where $s_1$ and $s_2$ are the slopes of the upstream and downstream linear profiles, respectively, and $x_0$ is the location of each characteristic at $t=0$.\\

In the proposed system, since signals propagate at a sound speed equal to the corresponding value of $\rho$ and only in the rightward direction, the characteristics in an $x$-$t$ diagram are straight lines along which $\rho$ is constant and always equals to its initial value at $t=0$. The function of each characteristic in upstream and downstream regions is as follows,
\begin{equation}
x_1=x_0+\rho_{0,1}t \;\;\; x_2=x_0+\rho_{0,2}t
\label{EqB2}
\end{equation}
Combining Eqs.~\ref{EqB1} and \ref{EqB2}, the constant value of $\rho$ along each characteristic can be expressed as
\begin{equation}
\rho_{0,1} = \frac{s_1 x_1 + \rho_1}{s_1 t + 1} \;\;\; \rho_{0,2} = \frac{s_2 x_2 + \rho_2}{s_2 t + 1} 
\label{EqB3}
\end{equation}
As shown in Fig.~\ref{FigB1}(b), the shock propagates along a trajectory where the characteristics from the upstream region intersect with those from the downstream region in $x$-$t$ space, determining the shock velocity. As obtained in Eq.~\ref{EqA3}, the speed of shock propagation at any time is equal to the average of the sound speeds corresponding to the two characteristics intersecting at this time. Since along each characteristic, the sound speed always equals to the initial value of $\rho$ at $t=0$, Eq.~\ref{EqA3} can be rewritten as
\begin{equation}
S=\frac{1}{2} \left(\rho_{0,1} + \rho_{0,2} \right)
\label{EqB4}
\end{equation}
Using Eq.~\ref{EqB3} to express $\rho_{0,1}$ and $\rho_{0,2}$ in Eq.~\ref{EqB4} and substituting $x_1$ and $x_2$ by the position of the shock, $x_{\!_\mathrm{S}}$, a first-order linear ordinary differential equation governing the propagation of the shock wave is obtained,
\begin{equation}
S = \frac{\mathrm{d} x_{\!_\mathrm{S}}}{\mathrm{d} t} = \frac{1}{2} \left( \frac{s_1 x_1 + \rho_1}{s_1 t + 1} + \frac{s_2 x_2 + \rho_2}{s_2 t + 1} \right)
\label{EqB5}
\end{equation}
with initial condition $x_{\!_\mathrm{S}}=0$ at $t=0$. By integrating Eq.~\ref{EqB5}, the position of the shock as a function of time can be solved,
\begin{eqnarray}\nonumber
x_{\!_\mathrm{S}}(t) &= \frac{1}{s_1-s_2} \big[ \rho_1 \left( \sqrt{s_1t+1} \sqrt{s_2t+1} - s_2t -1 \right) \\[0.8em]
&- \rho_2 \left( \sqrt{s_1t+1} \sqrt{s_2t+1} - s_1t -1 \right) \big]
\label{EqB6}
\end{eqnarray}
This solution is used in the main paper to track the trajectory of the shock waves.

\section*{Acknowledgements}
The authors would like to acknowledge a number of stimulating discussions with John Lee, Samuel Goroshin, Charles Kiyanda, Hoi Dick Ng, Matei Radulescu, and Vincent Tanguay that contributed to the development of this paper.

\bibliographystyle{physrev5}
\bibliography{detonation}

\end{document}